\documentclass[reprint,10pt,superscriptaddress,twocolumn,aps,prx]{revtex4-2}
\usepackage{amsmath}
\usepackage{amssymb}
\usepackage{graphicx}
\usepackage[colorlinks=true,citecolor=blue,linktocpage=true,linkcolor=blue,filecolor=blue,urlcolor=magenta]{hyperref}
\usepackage{physics}
\usepackage{bbm}
\usepackage{color}
\usepackage{dcolumn}
\usepackage{color}

\makeatletter
\def\maketitle{
\@author@finish
\title@column\titleblock@produce
\suppressfloats[t]}
\makeatother

\begin{document}

\preprint{APS/123-QED}

\title{Feasibility Study of 3D-Printed Micro Junction Array\\ for Ion Trap Quantum Processor}

\author{Kento Taniguchi}
 \affiliation{Department of Physics, University of California, Berkeley, Berkeley, California 94720, USA}
\affiliation{Challenge Institute for Quantum Computation, University of California, Berkeley, Berkeley, California 94720, USA}%

\author{Ke Sun}
 \affiliation{Department of Physics, University of California, Berkeley, Berkeley, California 94720, USA}
\affiliation{Challenge Institute for Quantum Computation, University of California, Berkeley, Berkeley, California 94720, USA}%

\author{Shuqi Xu}
 \affiliation{Department of Physics, University of California, Berkeley, Berkeley, California 94720, USA}
\affiliation{Challenge Institute for Quantum Computation, University of California, Berkeley, Berkeley, California 94720, USA}%

\author{Abhinav Parakh}
 \affiliation{Engineering Directorate, Lawrence Livermore National Laboratory, Livermore, California 94550, USA}%

\author{Xiaoxing Xia}
 \affiliation{Engineering Directorate, Lawrence Livermore National Laboratory, Livermore, California 94550, USA}%

\author{Michael Schecter}
 \affiliation{Quantinuum, 303 South Technology Court, Broomfield, Colorado 80021, USA}%

\author{Curtis Volin}
 \affiliation{Oxford Ionics, Oxford, OX5 1GN, UK}%

\author{Eric Hudson}
\affiliation{Department of Physics and Astronomy, University of California, Los Angeles, Los Angeles, CA 90095, USA}
\affiliation{Challenge Institute for Quantum Computation, University of California, Los Angeles, Los Angeles, CA, USA}
\affiliation{Center for Quantum Science and Engineering, University of California, Los Angeles, Los Angeles, CA, USA}%

\author{Hartmut H\"{a}ffner}
\email{hhaeffner@berkeley.edu}
\affiliation{Department of Physics, University of California, Berkeley, Berkeley, California 94720, USA}
\affiliation{Challenge Institute for Quantum Computation, University of California, Berkeley, Berkeley, California 94720, USA}%

\date{\today}

\begin{abstract}
We introduce an ion trap platform based on a 3D-printed micro-junction array, designed to implement quantum charge-coupled device (QCCD) architectures for large-scale quantum information processing (QIP). 
The integration of three-dimensionally structured micro Radio-Frequency (RF) electrodes above a surface-electrode trap enables flexible control of electric field profiles across both linear and junction regions. 
Through simulations, we demonstrate that the linear region exhibits deeper and more harmonic ion confinement with reduced RF drive power compared to conventional planar traps. 
Crucially, we identify a junction geometry that maintains uniform ion confinement during transport while substantially reducing the pseudopotential barrier. 
This reduction facilitates low-heating, high-fidelity transport of single- and multi-species ion crystals. 
Our results establish a viable route toward fault-tolerant quantum computing by enabling modular and scalable QCCD systems based on the state-of-the-art 3D-printing technologies.
\end{abstract}

\maketitle


\section{Introduction}
Scalable quantum computers promise transformative advances in science and technology \cite{Beverland-2022-resources-scaling-QC}, yet remain hindered by the challenge of combining high-fidelity quantum operations with system scalability. A promising strategy is to spatially separate quantum operations into distinct zones while coherently transferring quantum information between them. This modular approach was originally proposed for trapped-ion systems as the quantum charge-coupled device (QCCD) architecture \cite{wineland1998, Kielpinski2002}, and is now an essential building block of neutral atom platforms based on optical tweezer arrays \cite{Bluvstein2022}. 

Trapped ion systems have demonstrated exceptional capabilities for quantum information processing (QIP), including the record for the highest two-qubit gate fidelity \cite{Srinivas2021, Clark2021, löschnauer2024}, and the demonstration of QIP with reconfigurable $56$ ions \cite{Moses2023}.  In their QCCD architectures, multiple trapping zones are connected by junctions that allow dynamic qubit reconfiguration and arbitrary connectivity \cite{Mordini2025}. Therefore, coherent ion transport through junctions lies at the heart of the QCCD architecture and has been the subject of extensive study \cite{Blakestad2009, Delaney2024}. Recent experiments have achieved ion shuttling through junctions with sub-quantum motional excitation, both for single ions and multi-species ion crystals \cite{Burton2023}, marking a milestone toward modular, reconfigurable, and large-scale QIP.

Advances in microfabrication technology have enabled the realization of complex and highly optimized surface electrodes, forming the backbone of the advancements in QCCD architecture \cite{chiaverini2005, Seidelin2006}. However, the intrinsic two-dimensional manifold of planar electrodes imposes constraints on trap geometry and electric field control. Consequently, surface traps suffer from weak and anharmonic confinement compared to macroscopic three-dimensional traps \cite{Wesenberg2008, Xu2025, Nguyen2025}, resulting in reduced trap depth and enhanced sensitivity to stray electric field fluctuations \cite{Wang2011, Charles2012}. These effects reduce the trap frequency of an ion, limiting the speed and fidelity of key operations such as quantum logic gates, laser cooling, and the splitting and merging of ion crystals, which dominate the duty cycle of the QCCD architecture \cite{Home2006, Pino2021, Moses2023}. 

The limitations are particularly severe near junctions, where the electric field profile becomes highly intricate, inducing unwanted motional excitation during ion transportation. While ion transportation fidelity has been improved using the constant total confinement (CTC) paths \cite{Wright2013}, surface traps exhibit a significant mismatch between the CTC and the pseudopotential minimum paths \cite{Zhang2022}, hindering the perfect suppression of motional excitation and complicating the transport of multi-species ion crystals \cite{Burton2023}. Although there are ongoing efforts using stacked-wafer fabrication to mitigate those issues \cite{Ragg2019, Decaroli2021}, these designs remain topologically similar to surface traps and offer limited improvements and scalability \cite{See2013, Auchter2022}. A new trap platform with an alternative topology may be needed to overcome these fundamental limitations and advance a large-scale, high-performance QCCD architecture.

Three-dimensionally printed micro ion traps offer a promising solution to the limitations of traditional surface traps \cite{quinn2022, Biener2022-3D-trap, Xu2025}. Leveraging the high precision and geometric flexibility of 3D-printing technologies, these traps can provide strong confinement with low drive power while maintaining scalability and compact integrability \cite{Xu2025}. The resulting high trap frequencies with minimized power dissipation allow us to reduce cooling requirements, mitigate motional anomalous heating with increased ion-to-electrode separation, and simplify the thermal management essential for a large-scale ion array. Recent experiments have demonstrated the feasibility of this approach by achieving high-fidelity single- and two-qubit gates, in addition to the motional ground-state preparation using Doppler cooling alone, unveiling the potential of 3D-printed microtraps for a scalable quantum information processor \cite{Xu2025}.

In this work, we present a promising 3D-printed junction array design that leverages a fully three-dimensional manifold to overcome the limitations of conventional surface-electrode ion trap junctions. Simulations show that this approach may achieve near-perfect overlap between the CTC and the pseudopotential minimum paths, mitigating the pseudopotential gradient on the CTC path, and reducing the motional excitation near junctions by up to two orders of magnitude. This improvement, combined with the advantages of enhanced confinement inherent to 3D-printed micro ion traps, should enable faster quantum operations, coherent shuttling of single- and multi-species ion crystals, and efficient cooling, all essential for large-scale QCCD architectures.

\begin{figure}[t]
\includegraphics[width=8.4cm]{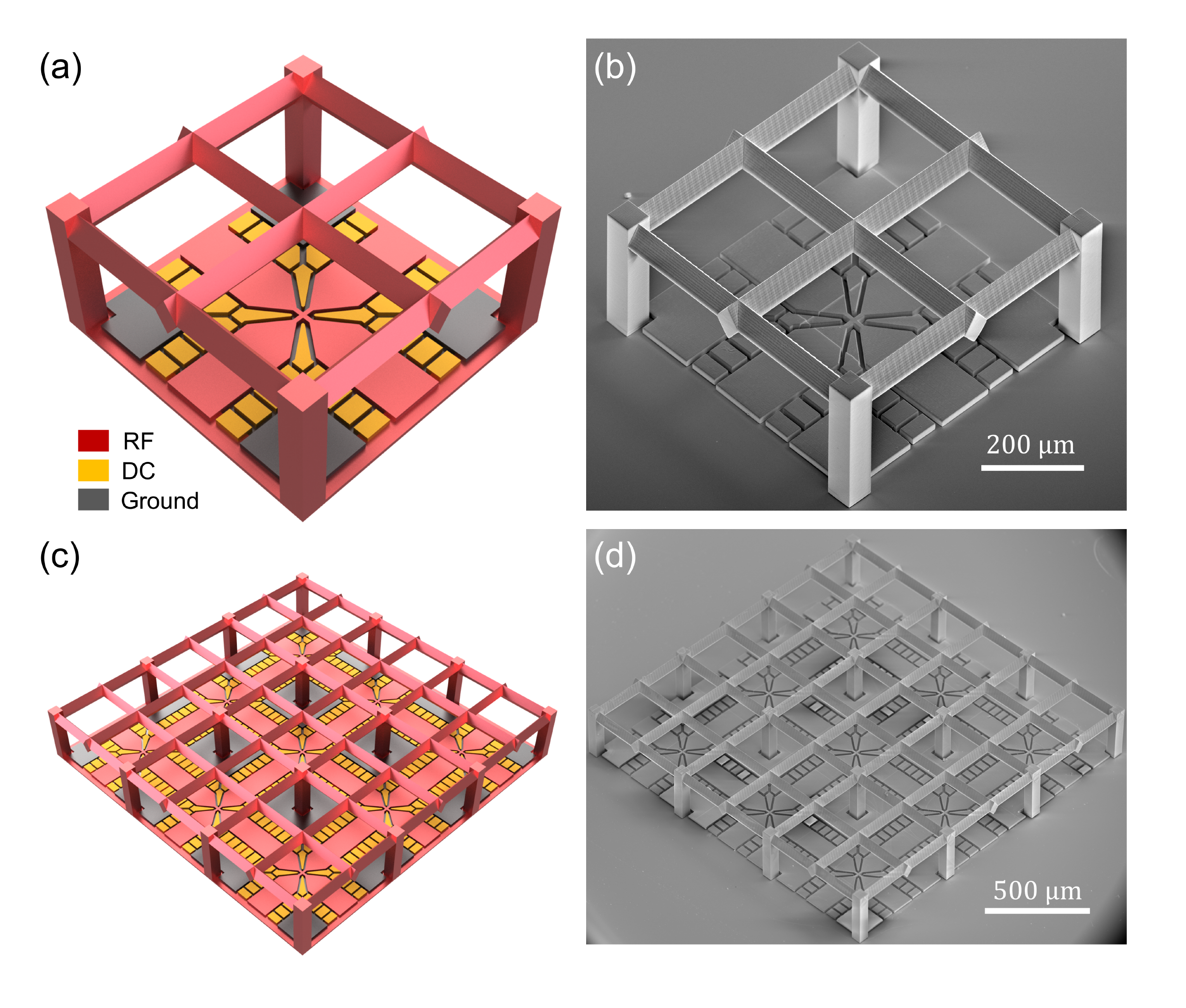}
\caption{\label{Fig1} Schematic configuration and the Scanning Electron Microscope (SEM) image of the 3D-printed micro X-junction array. (a) Schematic of a single X-junction unit within the array. The red, yellow, and gray regions correspond to the RF, DC, and ground electrodes, respectively. Across the entire device, the minimum RF-DC electrode spacing and RF electrode width are set to $10\,$\textmu{}m to prevent electrical shorts and ensure reliable wiring. A 3D-printed RF electrode is positioned $247\,$\textmu{}m above the surface with a diamond-shaped cross-section measuring $56\,$\textmu{}m horizontally and $82\,$\textmu{}m vertically. These structures are supported by identical diamond-shaped beams arranged in a grid and anchored on $56\,$\textmu{}m square pedestals, which are electrically connected to the surface RF electrodes. The overhanging electrode obstructs part of the ion’s emission toward the top imaging system, blocking $65.3\,\%$ of the field of view for a 0.3 NA lens and $28.6\,\%$ for a 0.6 NA lens. However, future designs may replace the solid top RF rail with a meshed structure, improving photon collection efficiency and simplifying metal coating. Additionally, in a well-designed QCCD architecture, these junctions can serve solely as gate zones, with ions shuttled to dedicated detection regions for readout. (b) The SEM image of the fabricated X-junction unit. (c) Overview of the $3\times 3$ junction array. The center-to-center distance between adjacent junctions is $600\,$\textmu{}m. (d) The SEM image of the fabricated $3\times 3$ junction array.}
\end{figure}

\section{3D-Printed Micro Junction Array}
The schematic configuration of the micro X-junction array we propose and the Scanning Electron Microscope (SEM) image of the fabricated unit cell are depicted in Fig.~\ref{Fig1}(a) and \ref{Fig1}(b). Radio-Frequency (RF) electric fields are applied to trap electrodes laid out on the center of the planar substrate and the 3D-printed structure, to generate the pseudopotential for ion confinement. Each micro X-junction unit occupies a $600\,$\textmu{}m$\times 600\,$\textmu{}m area and comprises four branches connected via a central X-junction. Ion loading, state preparation, measurement, and gate operations are performed at the branches, while the junction facilitates the transportation of ions between them. Surface RF electrodes have a uniform width of $120\,$\textmu{}m along the branches but are shaped intricately at the junction to maximize transport efficiency. Segmented DC electrodes flanking the RF electrodes generate a time-dependent potential for ion loading, gate operations, and ion shuttling. 

The 3D microstructures can be fabricated using commercially available two-photon lithography (Nanoscribe). In this process, a femtosecond laser of $780\, \mathrm{nm}$ wavelength is focused into a photoresist layer deposited on the pre-fabricated surface trap and scanned in three dimensions following the desired geometry. As a result of the cross-linking of the photoresist, the exposed regions polymerize, forming a solid 3D structure. After removing the unexposed resist with developer, the structure is coated with a thick Au or Al metal layer via electron-beam evaporation. This technique enables the formation of fully metal-coated, high-resolution, three-dimensional microstructures directly on the trap chip with sub-micron precision and substantial design flexibility \cite{Xia2019, Gao2020}. Figure~\ref{Fig1}(c) and \ref{Fig1}(d) show the schematic and the SEM image of the fabricated $3 \times 3$ 3D-printed micron X-junction. Using this method, 200 densely packed 3D-printed ion traps were also fabricated on a single chip \cite{Xu2025}, showing its compatibility with a scalable ion trap architecture. 

In the proposed architecture, the 3D structures are printed on a conventional surface trap chip, allowing seamless integration of control electronics and optical components on the underlying substrate with the approaches pursued with planar traps \cite{Mehta2016}. The support posts can also serve as shields for stray optical and electromagnetic fields between linear regions (the center of branches), suppressing crosstalk from neighboring trapping and optical units. Therefore, this approach offers a solution to miniaturize the integrated photonics QCCD systems further.

\begin{figure}[t]
\includegraphics[width=8.4cm]{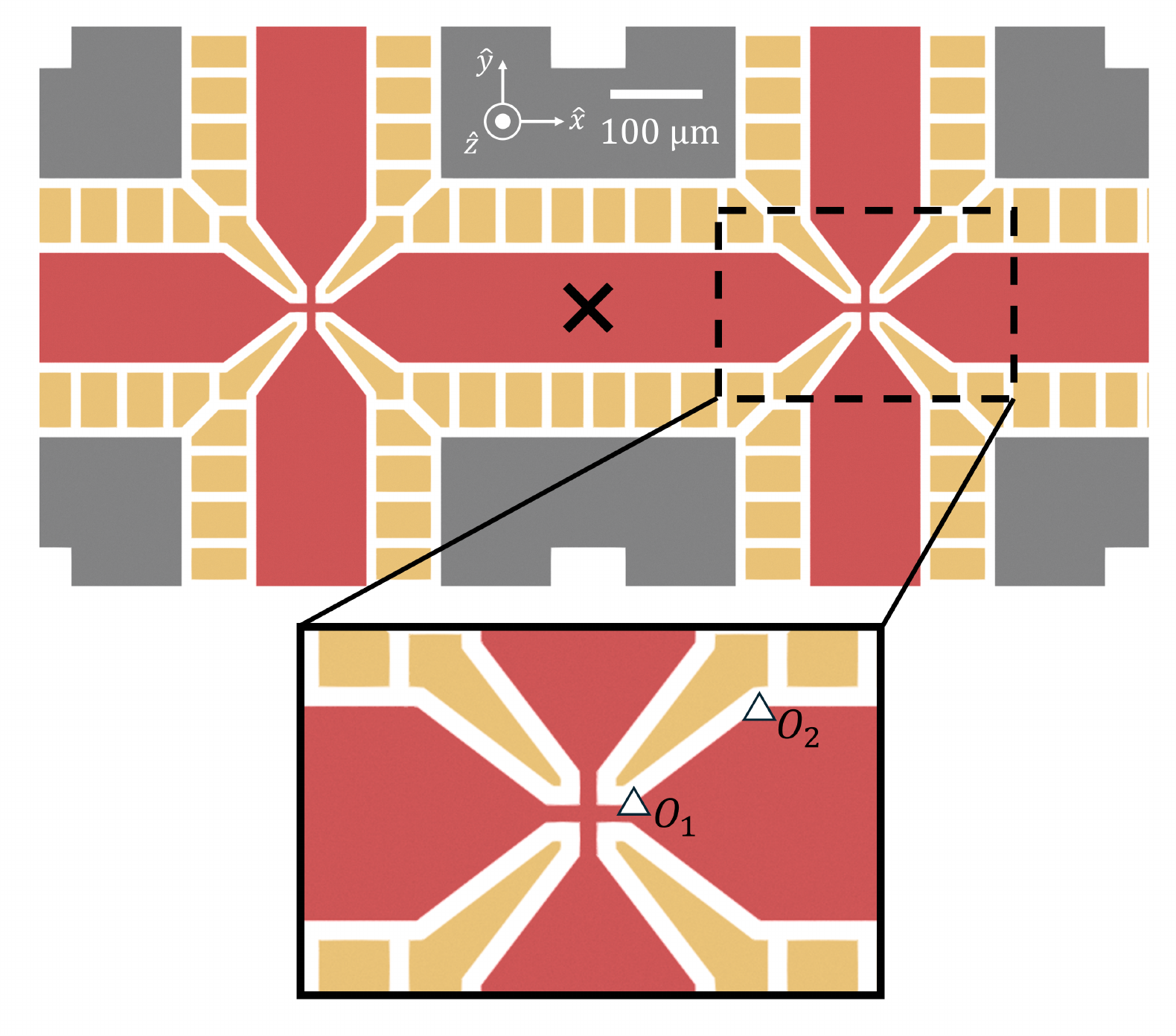}
\caption{\label{Fig2} The top view of the surface-electrode trap located beneath the 3D-printed RF electrodes. The black cross schematically denotes the linear trapping region. The inset details the geometry of the X-junction, where the positions of the RF electrode at $O_1$ and $O_2$ (indicated by triangles) are optimized to minimize the gradient of the pseudopotential barrier along the CTC path. This optimization suppresses RF noise-induced heating during each round trip.}
\end{figure}

\section{Linear Region}
The linear trapping region between junctions, as illustrated as a schematic black cross in Fig.~\ref{Fig2}, facilitates ion loading, state preparation, and readout, as well as single- and two-qubit gates. Therefore, the pseudopotential profile $\Phi_\mathrm{pp} (\vec{x})$ in this region is a critical factor in determining the overall performance of quantum operations. 

We employed the Finite Element Method (FEM) using COMSOL Multiphysics to simulate the electric field profile in the linear trapping region. To incorporate cross-talk effects from neighboring junctions, the simulation model included a configuration where the linear trapping region was connected by two X-junctions, as illustrated in Fig.~\ref{Fig2}. In the simulation, all DC and grounded electrodes were set to the ground potential, while an RF voltage was applied to the RF electrodes. From the resulting electric fields, we derived the pseudopotential for a $^{171}\mathrm{Yb}^{+}$ ion, assuming a drive frequency of $44.3 \, \mathrm{MHz}$ and a total confinement strength of $ C = \nabla^2 \Phi_\mathrm{pp} = 0.75 \times 10^{9} \, \mathrm{eV/m^2}$, consistent with the previous work \cite{Burton2023}.

\begin{figure}[t]
\includegraphics[width=8.4cm]{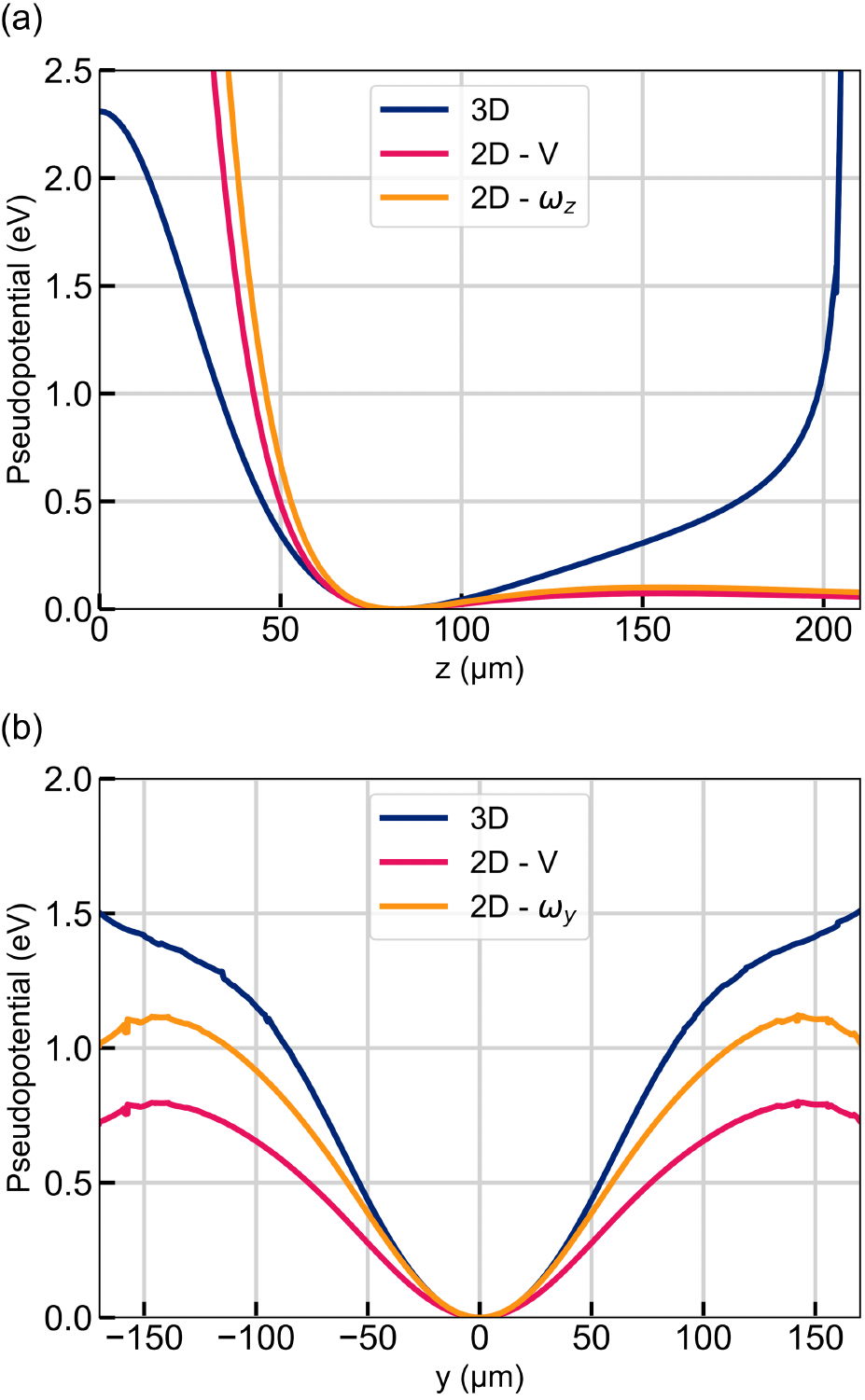}
\caption{\label{Fig3} Comparison of pseudopotential distributions at the linear trapping region. The pseudopotential generated by the 3D-printed micro X-junction array (blue) is compared with that of a conventional surface-electrode trap operated under two conditions: identical RF amplitude (red) and identical trap frequency (yellow). }
\end{figure}

The pseudopotential profile for the linear region of the 3D-printed micron junction array is presented in Fig.~\ref{Fig3}. For an RF amplitude of $190\, \mathrm{V}$, the 3D-printed trap generates a potential minimum $82.3\,$\textmu{}m above the surface, with a trap frequency of $\omega_z/2\pi = 2.32 \, \mathrm{MHz}$, the stability $q$ parameter of $q_z = 0.15$, and a pseudopotential trap depth of $2.3\, \mathrm{eV}$ along the $z$-axis, presented as a blue line in Fig.~\ref{Fig3}(a). To enable a direct comparison, we simulated a conventional surface-electrode trap with an RF width of $120 \,$\textmu{}m and an RF-RF electrodes separation of $71.5 \,$\textmu{}m, adjusted to yield the same trap height. Under identical RF voltage, the surface trap exhibited a trap frequency of $1.98 \, \mathrm{MHz}$ and a significantly shallower pseudopotential trap depth of $74\, \mathrm{meV}$, presented as a red line. This corresponds to a $17\,\%$ increase in trap frequency and a $31$-fold enhancement in pseudopotential trap depth for the 3D-printed trap. Moreover, even when the RF voltage for the surface trap is increased by $17\,\%$---corresponding to a $37\,\%$ increase in RF power---to match the 3D-printed trap's secular frequency, the resulting depth reaches only $100\,\mathrm{meV}$, depicted as a orange line. Therefore, the 3D-printed trap achieves a $23$-fold deeper pseudopotential while requiring $37\,\%$ less RF power. The pseudopotential along the $y$-axis, shown in Fig.~\ref{Fig3}(b), exhibits similar enhancements in trap frequency by $18\, \%$, confinement efficiency by $40\, \%$, and pseudopotential trap depth by $5$-fold. These results demonstrate that the 3D-printed structure provides a deeper confinement and a similar or slightly better trap frequency and efficiency compared to conventional surface-electrode traps.

The pseudopotential can be more precisely characterized by expanding it as
\begin{equation}
    q\Phi_\mathrm{pp} (x,y,z) = \frac{1}{2}m \omega_\mathrm{z}^2 \sum_{i=2}^{\infty} \sum_{j=2}^{\infty}\sum_{k=2}^{\infty}C_{ijk}x^i y^j z^k,
\end{equation}
where $q$ and $m$ are the charge and the mass of the trapped ion, $\omega_z$ is the trap frequency along the $z$-axis, and $C_{ijk}$ are normalized coefficients representing the strength of anharmonicity in the pseudopotential. As listed in Table.~\ref{table1}, the 3D-printed trap exhibits reduced anharmonic terms: the third- and fourth-order coefficients along the $z$-axis are smaller by factors of approximately $2$ and $4$, respectively, while in the $y$-direction, the reduction exceeds an order of magnitude compared to the surface-electrode trap. Anharmonicity in the trap potential leads to amplitude-dependent shifts in the trap frequency, described by \cite{Goldman2010}
\begin{equation}
    \omega_z (A_z) = \omega_z \left[1 + \sum_{k=2}^{\infty}\alpha_k A_z^k \right],
\end{equation}
where $A_z$ represents the amplitude of motion, and the coefficients $\alpha_k$ quantify the anharmonicity-induced frequency deviations along the $z$-axis, related to the potential expansion coefficients by
\begin{align}
    \alpha_2 &= - \frac{15(C_{003})^2}{16} + \frac{3 C_{004}}{4} \\
    \alpha_3 &= (C_{003}) \alpha_2.
\end{align}
These frequency shifts are particularly sensitive to perturbations in the trap potential minimum, which can arise from fluctuations in stray electric fields \cite{Xu2025}. 

\begin{table}[t]
    \caption{The anharmonicity coefficients of the pseudopotential generated by the 3D-printed micro trap and the conventional surface-electrode trap at the linear region.}
    \centering
    \begin{tabular}{rrrr} \hline \hline
         &  \multicolumn{1}{c}{3D}  & \multicolumn{1}{c}{2D} & \multicolumn{1}{c}{units} \\ \hline
        \multicolumn{1}{c}{$C_{200}$} & \multicolumn{1}{c}{$-1.1 \times 10^{-3}$} & \multicolumn{1}{c}{---} & \multicolumn{1}{c}{$1$} \\
        \multicolumn{1}{c}{$C_{201}$} & \multicolumn{1}{c}{\phantom{0} $5.2 \times 10^{-4}$} & \multicolumn{1}{c}{---} & \multicolumn{1}{c}{\textmu{}$\mathrm{m}^{-1}$
        } \\
        \multicolumn{1}{c}{$C_{400}$} & \multicolumn{1}{c}{\phantom{0} $9.1 \times 10^{-6}$} & \multicolumn{1}{c}{---} & \multicolumn{1}{c}{\textmu{}$\mathrm{m}^{-2}$
        } \\
        \multicolumn{1}{c}{$C_{020}$} & \multicolumn{1}{c}{$1$} & \multicolumn{1}{c}{$1$} & \multicolumn{1}{c}{$1$
        } \\
        \multicolumn{1}{c}{$C_{021}$} & \multicolumn{1}{c}{$-1.8 \times 10^{-2}$} & \multicolumn{1}{c}{$-3.9 \times 10^{-2}$} & \textmu{}$\mathrm{m}^{-1}$ \\
        \multicolumn{1}{c}{$C_{040}$} & \multicolumn{1}{c}{$-4.0 \times 10^{-5}$} & \multicolumn{1}{c}{$-4.9 \times 10^{-4}$} & \multicolumn{1}{c}{\textmu{}$\mathrm{m}^{-2}$} \\
        \multicolumn{1}{c}{$C_{002}$} & \multicolumn{1}{c}{$1$} & \multicolumn{1}{c}{$1$} & \multicolumn{1}{c}{$1$
        } \\
        \multicolumn{1}{c}{$C_{003}$} & $- 1.9 \times 10^{-2}$ & $- 3.9 \times 10^{-2}$ & \textmu{}$\mathrm{m}^{-1}$ \\
        \multicolumn{1}{c}{$C_{004}$} & $1.9 \times 10^{-4}$ & $7.4 \times 10^{-4}$& \textmu{}$\mathrm{m}^{-2}$ \\
        
     \hline \hline
    \label{table1}
    \end{tabular}
\end{table}

A high-fidelity two-qubit gate depends on precise matching of motional frequency, spin frequency, and external drives. Therefore, frequency shifts caused by trap anharmonicity critically impact gate performance \cite{Clark2021}. However, anharmonicity worsens relative to the trap frequency for smaller ion height because the ratio $(\partial^3\Phi_\mathrm{pp}/\partial z^3)/(\partial^2\Phi_\mathrm{pp}/\partial z^2)$ increases sensitively, and at least inversely, with ion-electrode distance. Thus, anharmonicity in trap potential limits the miniaturization of trap size, prohibiting the realization of a large-scale trapped ion system. Owing to its lower intrinsic anharmonicity, the 3D-printed trap demonstrates better resilience against such disturbances, offering improved frequency stability and miniaturizability essential for a scalable and high-fidelity QCCD architecture.

Finally, we evaluated the residual pseudopotential near the linear trapping region. Due to interference from surrounding junctions and 3D-printed structures, even slight deviations of the ion position from the linear trapping region could increase the local pseudopotential offset, thereby inducing excess micromotion along the shuttling direction ($x$-axis). Within the pseudopotential approximation, the micromotion can be expressed as \cite{Zhang2022}
\begin{equation}
    x(t) \approx - \sqrt{\frac{4q \Phi_\mathrm{pp}}{m \Omega_\mathrm{rf}^2}} \cos{(\Omega_\mathrm{rf}t)}
\end{equation}
where $\Phi_\mathrm{pp}$ is the residual pseudopotential and $\Omega_\mathrm{rf}$ the drive frequency of the trap. Our simulation revealed a residual pseudopotential of $5\times10^{-5} \, \mathrm{eV}$ over a total span of $20\,$\textmu{}m along the axial direction ($x$-axis) around the linear region, corresponding to a micromotion amplitude of $38 \, \mathrm{nm}$, compatible with previously reported values \cite{Blakestad2011}. These results confirm that the 3D-printed trap supports a sufficiently uniform linear trapping region without introducing significant micromotion.

\section{Junction Region}
A well-designed junction is crucial for ensuring reliable and coherent ion transport. Ideal transport paths are those that nullify the pseudopotential (RF electric fields) while maintaining stable mode frequencies and a large trap depth. However, enforcing the conditions of zero RF electric fields along paths that intersect at a junction results in the disappearance of the quadrupole term of the electric potential at the intersection point \cite{Wesenberg2009}. In the linear trap region discussed in the previous section, the quadrupole potential provides the restoring force necessary for ion confinement at a well-defined trap frequency, characterized by the total confinement $C=\nabla^2 \Phi_\mathrm{pp} \propto \sum_i \omega_i^2$, where the $\omega_i/2\pi$ is the $i$-th normal motional frequency of a single ion. However, this confinement weakens as the quadrupole component diminishes near the junction, leading to a reduced trap frequency and trap depth. This degradation enhances the ions' susceptibility to motional excitation by stray electric fields and non-adiabatic processes, thereby increasing the likelihood of motional excitation and ion loss.

An efficient strategy to mitigate the reduction of the trap frequency is to employ trajectories along which the total confinement remains constant, referred to as constant total confinement (CTC) paths \cite{Wright2013, Zhang2022}. Along these paths, higher-order anharmonic components of the electric potential compensate for the local reduction of the quadrupole potential near junctions, thereby sustaining the overall trapping strength. Experimental implementations of this method have demonstrated low motional excitation rates and robustness during ion shuttling across X-junctions \cite{Burton2023}. 

Despite the advantages of the CTC path, surface-electrode traps impose a challenge arising from their planar geometry: the CTC path often fails to align with the pseudopotential minimum path \cite{Zhang2022, Burton2023}. Generally, the CTC paths tend to pass closer to the electrode surface, where the influence of higher-order anharmonic potentials is stronger, enabling the compensation of reduced quadrupole confinement. In contrast, the pseudopotential minimum typically appears farther from the surface as the RF electric potential diminishes with increasing distance from the electrodes. This geometric mismatch results in a persistent pseudopotential barrier along the CTC path. As ions traverse this barrier, they are subjected to steep pseudopotential gradients, which can cause motional excitation in the presence of RF noise. The corresponding heating rate is quantitatively described by \cite{Blakestad2009, Blakestad2011}
\begin{equation}
    \dot{\bar{n}} = \frac{q^4}{16 m^3 \Omega_\mathrm{rf}^4 \hbar \omega_x }\left[ \frac{\partial}{\partial x} E^2 (x)\right]^2 \frac{S_{V_N}(\Omega_\mathrm{rf} + \omega_x)}{V_\mathrm{rf}^2},
    \label{heat}
\end{equation}
where $\omega_x$ is the angular frequency of the axial motion, $\partial E^2(x) /\partial x$ represents gradient of the pseudopotential barrier along the shuttling direction, $S_{V_N}(\Omega_\mathrm{rf} + \omega_x)$ is the voltage-noise spectral density at RF sideband frequency, and $V_\mathrm{rf}$ is the amplitude of the RF potential. 

While increasing the shuttling speed can reduce the exposure time to the barrier and partially suppress heating, it also introduces a competing effect: resonance between the ion's motion and digital-to-analog converter (DAC)-induced noise, particularly when the update rate of the DAC becomes commensurate with the ion's trap frequency \cite{Blakestad2011}. This trade-off poses a practical constraint on the transport speed and the motional excitation. An alternative approach involves sympathetic cooling using co-trapped auxiliary ion species \cite{Burton2023, Delaney2024}. Although effective in removing excess motional energy, this method introduces additional complexity. Apart from the additional lengthy cooling process, species-dependent pseudopotentials lead to variations in confinement curvature, which can cause differential motional excitation and even ion loss. Furthermore, the position of the potential minimum shifts depending on the ion species, complicating the coherent transport of multi-species ion crystals. 

\begin{figure}[t]
\includegraphics[width=8.4cm]{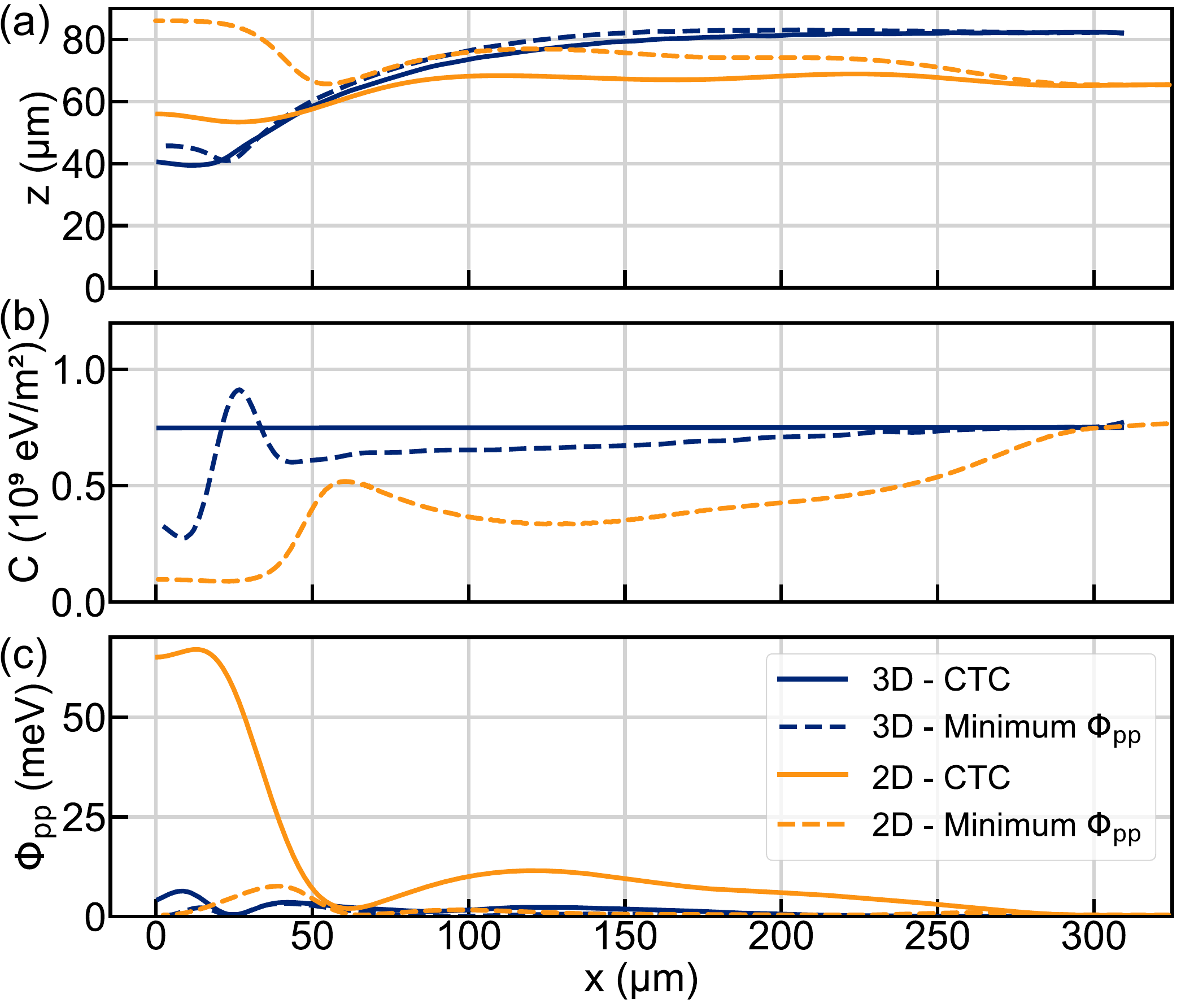}
\caption{\label{Fig4} Pseudopotential analysis between the junction and linear region along the pseudopotential minimum path (solid lines) and the CTC path (dashed lines). The solid and dashed blue lines are those for the present 3D-printed microjunction array, while the solid and dashed yellow lines are for the design from the previous work \cite{Burton2023}. (a) Trap height above the electrode surface along each path. The two trajectories are nearly identical, indicating close spatial overlap. (b) Total confinement along each path. The confinement degradation along the pseudopotential minimum path is mitigated due to its alignment with the CTC path. We omitted the CTC path from the previous design because it is identical to the one for the present design (dashed yellow line). (c) Pseudopotential experienced by a $^{171}\mathrm{Yb}^{+}$ ion along both paths. The pseudopotential barrier along the CTC path remains below $7\, \mathrm{meV}$.}
\end{figure}

We investigated the pseudopotential minimum and CTC paths for the 3D-printed micro X-junction. To extract the pseudopotential minimum path, we identified the $z$-coordinate corresponding to the local minimum of the pseudopotential at each point along the $x$-axis, with $y$ fixed at zero. For the CTC path, the total confinement was set to $C = 0.75\times10^{9} \, \mathrm{eV/m^2}$, consistent with the value in the linear region and in agreement with the previous research \cite{Burton2023}. Figure~\ref{Fig4} illustrates the variation in trap height, total confinement, and pseudopotential along both paths, extending from the junction center ($x=0\,$\textmu{}m) to the linear region ($x=300 \,$\textmu{}m).

Figure~\ref{Fig4}(a) shows the trap height along the pseudopotential minimum and the CTC paths for both the present 3D-printed microjunction array and the previous design reported in \cite{Burton2023}. The solid and dashed blue lines correspond to the pseudopotential minimum and CTC paths in the present design. In contrast, the solid and dashed yellow lines represent the same paths from the previous work \cite{Burton2023}. The two paths deviated significantly near the junction in the X-junction design of the prior work \cite{Burton2023}. However, the current design exhibits nearly perfect alignment throughout the region of interest. This alignment is attributed to the electric fields generated by the 3D-printed RF electrodes, which deflect the pseudopotential minimum path toward the surface near the junction, reducing its height to approximately $40\,$\textmu{}m. The effect of this alignment can be seen in the total confinement plotted in Fig.~\ref{Fig4}(b). In previous work, the confinement near the junction decreased to $1/8$ of its value in the linear region \cite{Burton2023}. In contrast, the present design mitigates this reduction to approximately $1/2$. The observed improvement is attributable to the nearly perfect overlap between the pseudopotential minimum and the CTC paths, the latter maintaining the constant total confinement along its trajectory. Moreover, this alignment makes the pseudopotential barrier along the CTC path comparable to the minimum path as shown in Fig.~\ref{Fig4}(c), reducing it nearly one order of magnitude relative to earlier surface-electrode trap designs \cite{Burton2023}. 

\begin{figure}[t]
\includegraphics[width=8.4cm]{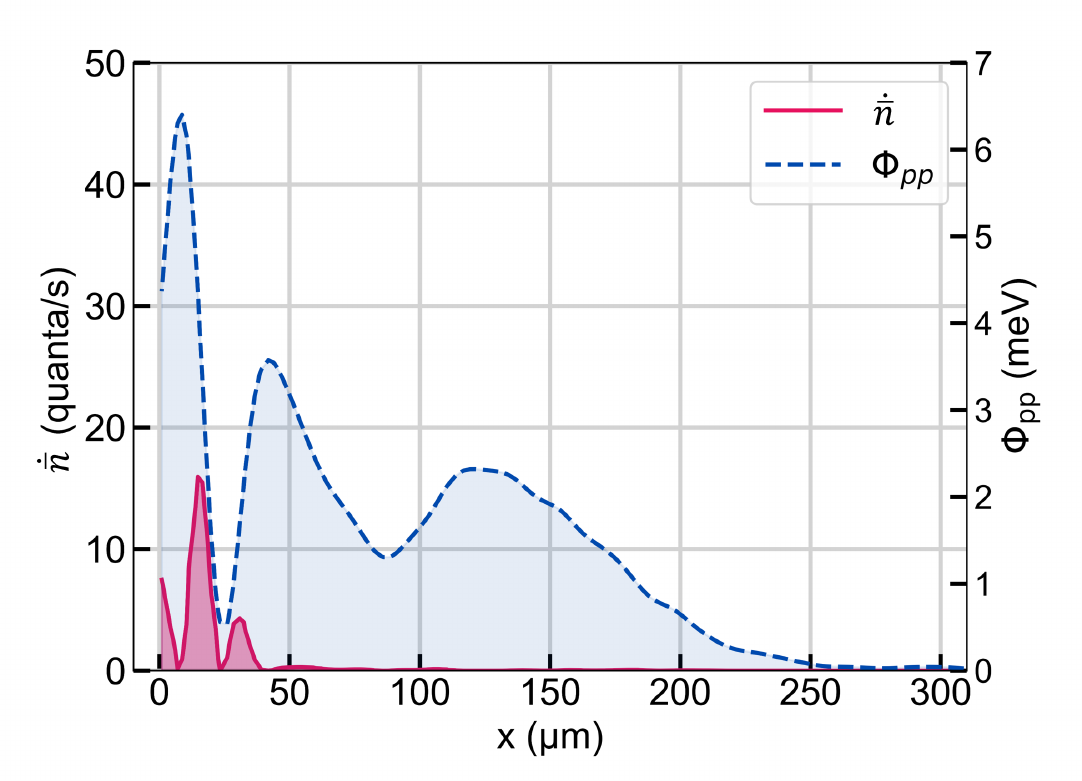}
\caption{\label{Fig5} Heating rate distribution $\dot{\bar{n}}$ (solid red) and pseudopotential profile (dashed blue) along the CTC path between the junction and linear region. The heating rate is evaluated at an axial trap frequency of $\omega_x/2\pi = 1.13\, \mathrm{MHz}$ with an RF noise level of $-178\,\mathrm{dBc}$. It is proportional to the square of the pseudopotential gradient and exhibits strong dependence on the steepness of the pseudopotential barrier.}
\end{figure}

We evaluated motion heating along the trap axis arising from RF noise at the pseudopotential barrier along the CTC path. Using an axial secular frequency of $\omega_x/2\pi = 1.13\, \mathrm{MHz}$ and an RF noise level of $-178\,\mathrm{dBc}$--- parameters consistent with \cite{Burton2023}---we calculated the position-dependent heating rate using the Eq.~(\ref{heat}), and overlaid it with the pseudopotential profile on the CTC path, as shown in Fig.~\ref{Fig5}. The analysis reveals that heating is predominantly localized near the junction, where symmetry breaking induces large pseudopotential barriers. Despite this concentration, the maximum heating rate remains below $16 \,\mathrm{quanta/s}$. It is useful to compare it to the heating expected from surface noises, because the ion-electrode distance along the CTC path near the junction is reduced to two-thirds of that in previous work \cite{Burton2023}, which could result in up to a fivefold increase in surface-noise induced heating \cite{Brownnutt2015}. At cryogenic temperatures, a heating rate of $\dot{n}_\mathrm{Sr} = 50 \, \mathrm{quanta/s}$ was reported for a $^{88}\mathrm{Sr}^{+}$ ion with an axial trap frequency of $\omega_\mathrm{Sr}/2\pi = 850\,\mathrm{kHz}$ and and ion-electrode distance of $d_\mathrm{Sr} = 38\,$\textmu{}m \cite{Sedlacek2018}. Extrapolating this to the present parameters of $\omega_\mathrm{Yb}/2\pi = 1.13\,\mathrm{MHz}$ and $d_\mathrm{Yb} = 40\,$\textmu{}m for a $^{171}\mathrm{Yb}^{+}$ ion, the expected surface-noise-induced heating rate, $\dot{n}_\mathrm{Yb}$, is given by \cite{Yu2022}
\begin{equation}
    \dot{n}_\mathrm{Yb} = \frac{m_\mathrm{Sr}}{m_\mathrm{Yb}} \left( \frac{\omega_\mathrm{Sr}}{\omega_\mathrm{Yb}}\right)^{1+\gamma} \left(\frac{d_\mathrm{Sr}}{
    d_\mathrm{Yb}} \right)^{\beta} \dot{n}_\mathrm{Sr},
\end{equation}
where $m_\mathrm{Sr}$ and $m_\mathrm{Yb}$ are the mass of $^{88}\mathrm{Sr}^{+}$ and  $^{171}\mathrm{Yb}^{+}$ ions, respectively, and $\gamma = 1.3$ and $\beta = 3.9$ are experimentally determined scaling exponents from \cite{Sedlacek2018}. This yields an estimated heating rate of $\dot{n}_\mathrm{Yb} = 11 \, \mathrm{quanta/s}$. Thus, the surface noise-induced heating is compatible with that induced by RF noise at cryogenic temperatures, even considering the reduced ion-electrode distance near the junction.

The total motional excitation during ion transport across junctions was quantified by integrating the position-dependent heating rate $\dot{\bar{n}}$ divided by the shuttling speed $v_x$ along the transport. As a normalized benchmark, we evaluated the excitation accumulated over a single round trip, defined as $\dot{\bar{n}}_\mathrm{tot} = \oint\dot{\bar{n}}/v_x dx$ where the integral path lies along the CTC path which goes from one linear trapping region across the junction to the opposite region and returning to the original position. Assuming a constant shuttling speed of $v_x = 4\,\mathrm{m/s}$--- corresponding to the average speed reported in \cite{Burton2023}---the total excitation was found to be $\dot{\bar{n}}_\mathrm{tot} =0.00019$ quanta per round-trip. This value is two orders of magnitude lower than the minimum heating observed in a state-of-the-art shuttling experiment \cite{Burton2023}, which was accurately predicted by the theoretical model employed in this study. This demonstrates the significant advantage of the 3D-printed micro X-junction over the conventional surface-electrode junctions.

\section{Conclusion}
This study introduces a novel QCCD architecture design based on a 3D-printed micro junction array, offering a fundamentally new approach to ion transport in scalable quantum systems. By exploiting the sub-micrometer design flexibility of 3D-printing technologies, the proposed architecture reduces excess motional excitation near junctions, which has been an enduring challenge in conventional surface-electrode traps. Beyond this key advancement, the architecture delivers multiple performance enhancements, including improved harmonicity and deeper trapping potentials, achieved without compromising trap frequency, scalability, and compatibility with integrated photonics. These advantages could enable faster, more reliable, and flexible quantum operations, providing a promising pathway toward realizing scalable, fault-tolerant quantum computing.

\section{Acknowledgments}
The authors thank Juergen Biener, Kristin Beck, Andrew Jayich, Boerge Hemmerling, and Karan Metha for the fruitful discussion. This research was supported by the NSF NQVL (Project No. 2435382). AP and XX's contributions to this work were performed under the auspices of the U.S. Department of Energy by Lawrence Livermore
National Laboratory under Contract DE-AC52-07NA27344 and was supported by the LLNL-LDRD Program under Project No. 23-ERD-021. LLNL-JRNL-2010042.

\bibliography{apssamp}
@article{haeffner2008,
title = {Quantum computing with trapped ions},
journal = {Physics Reports},
volume = {469},
number = {4},
pages = {155-203},
year = {2008},
issn = {0370-1573},
author = {H. Häffner and C.F. Roos and R. Blatt},
keywords = {Quantum computing and information, Entanglement, Ion traps},
}

@article{Kielpinski2002,
author={Kielpinski, D.
and Monroe, C.
and Wineland, D. J.},
title={Architecture for a large-scale ion-trap quantum computer},
journal={Nature},
year={2002},
month={6},
day={01},
volume={417},
number={6890},
pages={709-711},
}

@article{Bluvstein2022,
author={Bluvstein, Dolev
and Levine, Harry
and Semeghini, Giulia
and Wang, Tout T.
and Ebadi, Sepehr
and Kalinowski, Marcin
and Keesling, Alexander
and Maskara, Nishad
and Pichler, Hannes
and Greiner, Markus
and Vuleti{\'{c}}, Vladan
and Lukin, Mikhail D.},
title={A quantum processor based on coherent transport of entangled atom arrays},
journal={Nature},
year={2022},
month={4},
day={01},
volume={604},
number={7906},
pages={451-456},
}

@article{Srinivas2021,
author={Srinivas, R.
and Burd, S. C.
and Knaack, H. M.
and Sutherland, R. T.
and Kwiatkowski, A.
and Glancy, S.
and Knill, E.
and Wineland, D. J.
and Leibfried, D.
and Wilson, A. C.
and Allcock, D. T. C.
and Slichter, D. H.},
title={High-fidelity laser-free universal control of trapped ion qubits},
journal={Nature},
year={2021},
month={9},
day={01},
volume={597},
number={7875},
pages={209-213}
}

@article{Clark2021,
  title = {High-Fidelity Bell-State Preparation with $^{40}{\mathrm{Ca}}^{+}$ Optical Qubits},
  author = {Clark, Craig R. and Tinkey, Holly N. and Sawyer, Brian C. and Meier, Adam M. and Burkhardt, Karl A. and Seck, Christopher M. and Shappert, Christopher M. and Guise, Nicholas D. and Volin, Curtis E. and Fallek, Spencer D. and Hayden, Harley T. and Rellergert, Wade G. and Brown, Kenton R.},
  journal = {Phys. Rev. Lett.},
  volume = {127},
  issue = {13},
  pages = {130505},
  numpages = {5},
  year = {2021},
  month = {9},
  publisher = {American Physical Society}
}

@misc{löschnauer2024,
      title={Scalable, high-fidelity all-electronic control of trapped-ion qubits}, 
      author={C. M. Löschnauer and J. Mosca Toba and A. C. Hughes and S. A. King and M. A. Weber and R. Srinivas and R. Matt and R. Nourshargh and D. T. C. Allcock and C. J. Ballance and C. Matthiesen and M. Malinowski and T. P. Harty},
      year={2024},
      eprint={2407.07694},
      archivePrefix={arXiv},
      primaryClass={quant-ph},
      url={https://arxiv.org/abs/2407.07694}, 
}

@article{Moses2023,
  title = {A Race-Track Trapped-Ion Quantum Processor},
  author = {Moses, S. A. and Baldwin, C. H. and Allman, M. S. and Ancona, R. and Ascarrunz, L. and Barnes, C. and Bartolotta, J. and Bjork, B. and Blanchard, P. and Bohn, M. and Bohnet, J. G. and Brown, N. C. and Burdick, N. Q. and Burton, W. C. and Campbell, S. L. and Campora, J. P. and Carron, C. and Chambers, J. and Chan, J. W. and Chen, Y. H. and Chernoguzov, A. and Chertkov, E. and Colina, J. and Curtis, J. P. and Daniel, R. and DeCross, M. and Deen, D. and Delaney, C. and Dreiling, J. M. and Ertsgaard, C. T. and Esposito, J. and Estey, B. and Fabrikant, M. and Figgatt, C. and Foltz, C. and Foss-Feig, M. and Francois, D. and Gaebler, J. P. and Gatterman, T. M. and Gilbreth, C. N. and Giles, J. and Glynn, E. and Hall, A. and Hankin, A. M. and Hansen, A. and Hayes, D. and Higashi, B. and Hoffman, I. M. and Horning, B. and Hout, J. J. and Jacobs, R. and Johansen, J. and Jones, L. and Karcz, J. and Klein, T. and Lauria, P. and Lee, P. and Liefer, D. and Lu, S. T. and Lucchetti, D. and Lytle, C. and Malm, A. and Matheny, M. and Mathewson, B. and Mayer, K. and Miller, D. B. and Mills, M. and Neyenhuis, B. and Nugent, L. and Olson, S. and Parks, J. and Price, G. N. and Price, Z. and Pugh, M. and Ransford, A. and Reed, A. P. and Roman, C. and Rowe, M. and Ryan-Anderson, C. and Sanders, S. and Sedlacek, J. and Shevchuk, P. and Siegfried, P. and Skripka, T. and Spaun, B. and Sprenkle, R. T. and Stutz, R. P. and Swallows, M. and Tobey, R. I. and Tran, A. and Tran, T. and Vogt, E. and Volin, C. and Walker, J. and Zolot, A. M. and Pino, J. M.},
  journal = {Phys. Rev. X},
  volume = {13},
  issue = {4},
  pages = {041052},
  numpages = {25},
  year = {2023},
  month = {12},
  publisher = {American Physical Society}
}

@article{Mordini2025,
  title = {Multizone Trapped-Ion Qubit Control in an Integrated Photonics QCCD Device},
  author = {Mordini, Carmelo and Ricci Vasquez, Alfredo and Motohashi, Yuto and M\"uller, Mose and Malinowski, Maciej and Zhang, Chi and Mehta, Karan K. and Kienzler, Daniel and Home, Jonathan P.},
  journal = {Phys. Rev. X},
  volume = {15},
  issue = {1},
  pages = {011040},
  numpages = {17},
  year = {2025},
  month = {2},
  publisher = {American Physical Society}
}

@article{Delaney2024,
  title = {Scalable Multispecies Ion Transport in a Grid-Based Surface-Electrode Trap},
  author = {Delaney, Robert D. and Sletten, Lucas R. and Cich, Matthew J. and Estey, Brian and Fabrikant, Maya I. and Hayes, David and Hoffman, Ian M. and Hostetter, James and Langer, Christopher and Moses, Steven A. and Perry, Abigail R. and Peterson, Timothy A. and Schaffer, Andrew and Volin, Curtis and Vittorini, Grahame and Burton, William Cody},
  journal = {Phys. Rev. X},
  volume = {14},
  issue = {4},
  pages = {041028},
  numpages = {12},
  year = {2024},
  month = {11},
  publisher = {American Physical Society}
}

@article{Blakestad2009,
  title = {High-Fidelity Transport of Trapped-Ion Qubits through an $\mathbf{X}$-Junction Trap Array},
  author = {Blakestad, R. B. and Ospelkaus, C. and VanDevender, A. P. and Amini, J. M. and Britton, J. and Leibfried, D. and Wineland, D. J.},
  journal = {Phys. Rev. Lett.},
  volume = {102},
  issue = {15},
  pages = {153002},
  numpages = {4},
  year = {2009},
  month = {4},
  publisher = {American Physical Society}
}

@article{Burton2023,
  title = {Transport of Multispecies Ion Crystals through a Junction in a Radio-Frequency Paul Trap},
  author = {Burton, William Cody and Estey, Brian and Hoffman, Ian M. and Perry, Abigail R. and Volin, Curtis and Price, Gabriel},
  journal = {Phys. Rev. Lett.},
  volume = {130},
  issue = {17},
  pages = {173202},
  numpages = {6},
  year = {2023},
  month = {4},
  publisher = {American Physical Society}
}

@misc{chiaverini2005,
      title={Surface-Electrode Architecture for Ion-Trap Quantum Information Processing}, 
      author={J. Chiaverini and R. B. Blakestad and J. Britton and J. D. Jost and C. Langer and D. Leibfried and R. Ozeri and D. J. Wineland},
      year={2005},
      eprint={quant-ph/0501147},
      archivePrefix={arXiv},
      primaryClass={quant-ph},
      url={https://arxiv.org/abs/quant-ph/0501147}, 
}

@article{Seidelin2006,
  title = {Microfabricated Surface-Electrode Ion Trap for Scalable Quantum Information Processing},
  author = {Seidelin, S. and Chiaverini, J. and Reichle, R. and Bollinger, J. J. and Leibfried, D. and Britton, J. and Wesenberg, J. H. and Blakestad, R. B. and Epstein, R. J. and Hume, D. B. and Itano, W. M. and Jost, J. D. and Langer, C. and Ozeri, R. and Shiga, N. and Wineland, D. J.},
  journal = {Phys. Rev. Lett.},
  volume = {96},
  issue = {25},
  pages = {253003},
  numpages = {4},
  year = {2006},
  month = {6},
  publisher = {American Physical Society}
}

@article{Wesenberg2008,
  title = {Electrostatics of surface-electrode ion traps},
  author = {Wesenberg, J. H.},
  journal = {Phys. Rev. A},
  volume = {78},
  issue = {6},
  pages = {063410},
  numpages = {12},
  year = {2008},
  month = {12},
  publisher = {American Physical Society}
}

@misc{Nguyen2025,
      title={The Effect of Trap Design on the Scalability of Trapped-Ion Quantum Technologies}, 
      author={Le Minh Anh Nguyen and Brant Bowers and Sara Mouradian},
      year={2025},
      eprint={2503.00218},
      archivePrefix={arXiv},
      primaryClass={quant-ph},
      url={https://arxiv.org/abs/2503.00218}, 
}

@article{Wang2011,
    author = {Wang, Shannon X. and Hao Low, Guang and Lachenmyer, Nathan S. and Ge, Yufei and Herskind, Peter F. and Chuang, Isaac L.},
    title = {Laser-induced charging of microfabricated ion traps},
    journal = {Journal of Applied Physics},
    volume = {110},
    number = {10},
    pages = {104901},
    year = {2011},
    month = {11}
}

@article{Charles2012,
year = {2012},
month = {7},
publisher = {IOP Publishing},
volume = {14},
number = {7},
pages = {073012},
author = {Charles Doret, S and Amini, Jason M and Wright, Kenneth and Volin, Curtis and Killian, Tyler and Ozakin, Arkadas and Denison, Douglas and Hayden, Harley and Pai, C-S and Slusher, Richart E and Harter, Alexa W},
title = {Controlling trapping potentials and stray electric fields in a microfabricated ion trap through design and compensation},
journal = {New Journal of Physics}
}

@article{Home2006,
year = {2006},
month = {9},
publisher = {},
volume = {8},
number = {9},
pages = {188},
author = {Home, J P and McDonnell, M J and Lucas, D M and Imreh, G and Keitch, B C and Szwer, D J and Thomas, N R and Webster, S C and Stacey, D N and Steane, A M},
title = {Deterministic entanglement and tomography of ion–spin qubits},
journal = {New Journal of Physics}
}

@article{Pino2021,
author={Pino, J. M.
and Dreiling, J. M.
and Figgatt, C.
and Gaebler, J. P.
and Moses, S. A.
and Allman, M. S.
and Baldwin, C. H.
and Foss-Feig, M.
and Hayes, D.
and Mayer, K.
and Ryan-Anderson, C.
and Neyenhuis, B.},
title={Demonstration of the trapped-ion quantum CCD computer architecture},
journal={Nature},
year={2021},
month={4},
day={01},
volume={592},
number={7853},
pages={209-213}
}

@article{Zhang2022,
year = {2022},
month = {7},
publisher = {IOP Publishing},
volume = {24},
number = {7},
pages = {073030},
author = {Zhang, Chi and Mehta, Karan K and Home, Jonathan P},
title = {Optimization and implementation of a surface-electrode ion trap junction},
journal = {New Journal of Physics}
}

@article{Auchter2022,
year = {2022},
month = {5},
publisher = {IOP Publishing},
volume = {7},
number = {3},
pages = {035015},
author = {Auchter, S and Axline, C and Decaroli, C and Valentini, M and Purwin, L and Oswald, R and Matt, R and Aschauer, E and Colombe, Y and Holz, P and Monz, T and Blatt, R and Schindler, P and Rössler, C and Home, J},
title = {Industrially microfabricated ion trap with 1 eV trap depth},
journal = {Quantum Science and Technology}
}

@article{See2013,
  author={See, Patrick and Wilpers, Guido and Gill, Patrick and Sinclair, Alastair G.},
  journal={Journal of Microelectromechanical Systems}, 
  title={Fabrication of a Monolithic Array of Three Dimensional Si-based Ion Traps}, 
  year={2013},
  volume={22},
  number={5},
  pages={1180-1189},
  keywords={Electrodes;Geometry;Radio frequency;Electric potential;Gold;Three-dimensional displays;Microfabrication;Information processing;Ions;Surface treatment;Ion trap;microelectromechanical systems;quantum information processing;semiconductor devices}}

@article{Decaroli2021,
year = {2021},
month = {7},
publisher = {IOP Publishing},
volume = {6},
number = {4},
pages = {044001},
author = {Decaroli, Chiara and Matt, Roland and Oswald, Robin and Axline, Christopher and Ernzer, Maryse and Flannery, Jeremy and Ragg, Simon and Home, Jonathan P},
title = {Design, fabrication and characterization of a micro-fabricated stacked-wafer segmented ion trap with two X-junctions},
journal = {Quantum Science and Technology}
}

@article{Ragg2019,
    author = {Ragg, Simon and Decaroli, Chiara and Lutz, Thomas and Home, Jonathan P.},
    title = {Segmented ion-trap fabrication using high precision stacked wafers},
    journal = {Review of Scientific Instruments},
    volume = {90},
    number = {10},
    pages = {103203},
    year = {2019},
    month = {10}
}

@misc{Xu2023,
      title={3D-Printed Micro Ion Trap Technology for Scalable Quantum Information Processing}, 
      author={Shuqi Xu and Xiaoxing Xia and Qian Yu and Sumanta Khan and Eli Megidish and Bingran You and Boerge Hemmerling and Andrew Jayich and Juergen Biener and Hartmut Häffner},
      year={2023},
      eprint={2310.00595},
      archivePrefix={arXiv},
      primaryClass={quant-ph},
      url={https://arxiv.org/abs/2310.00595}, 
}

@misc{quinn2022,
      title={Geometries and fabrication methods for 3D printing ion traps}, 
      author={A. Quinn and M. Brown and T. J. Gardner and D. T. C. Allcock},
      year={2022},
      eprint={2205.15892},
      archivePrefix={arXiv},
      primaryClass={quant-ph},
      url={https://arxiv.org/abs/2205.15892}, 
}

@article{Xia2019,
author={Xia, Xiaoxing
and Afshar, Arman
and Yang, Heng
and Portela, Carlos M.
and Kochmann, Dennis M.
and Di Leo, Claudio V.
and Greer, Julia R.},
title={Electrochemically reconfigurable architected materials},
journal={Nature},
year={2019},
month={9},
day={01},
volume={573},
number={7773},
pages={205-213}
}

@article{Gao2020,
author = {Gao, Hongwei and Chen, George F. R. and Xing, Peng and Choi, Ju Won and Low, Hong Yee and Tan, Dawn T. H.},
title = {High-Resolution 3D Printed Photonic Waveguide Devices},
journal = {Advanced Optical Materials},
volume = {8},
number = {18},
pages = {2000613},
year = {2020}
}

@article{Blakestad2011,
  title = {Near-ground-state transport of trapped-ion qubits through a multidimensional array},
  author = {Blakestad, R. B and Ospelkaus, C. and VanDevender, A. P and Wesenberg, J. H and Biercuk, M. J and Leibfried, D. and Wineland, D. J},
  journal = {Phys. Rev. A},
  volume = {84},
  issue = {3},
  pages = {032314},
  numpages = {14},
  year = {2011},
  month = {9},
  publisher = {American Physical Society}
}

@article{Goldman2010,
  title = {Optimized planar Penning traps for quantum-information studies},
  author = {Goldman, J. and Gabrielse, G.},
  journal = {Phys. Rev. A},
  volume = {81},
  issue = {5},
  pages = {052335},
  numpages = {25},
  year = {2010},
  month = {5},
  publisher = {American Physical Society}
}

@article{Wesenberg2009,
  title = {Ideal intersections for radio-frequency trap networks},
  author = {Wesenberg, J. H.},
  journal = {Phys. Rev. A},
  volume = {79},
  issue = {1},
  pages = {013416},
  numpages = {7},
  year = {2009},
  month = {1},
  publisher = {American Physical Society}
}

@article{Hite2012,
  title = {100-Fold Reduction of Electric-Field Noise in an Ion Trap Cleaned with In Situ Argon-Ion-Beam Bombardment},
  author = {Hite, D. A. and Colombe, Y. and Wilson, A. C. and Brown, K. R. and Warring, U. and J\"ordens, R. and Jost, J. D. and McKay, K. S. and Pappas, D. P. and Leibfried, D. and Wineland, D. J.},
  journal = {Phys. Rev. Lett.},
  volume = {109},
  issue = {10},
  pages = {103001},
  numpages = {5},
  year = {2012},
  month = {9},
  publisher = {American Physical Society}
}

@article{Sedlacek2018,
  title = {Distance scaling of electric-field noise in a surface-electrode ion trap},
  author = {Sedlacek, J. A. and Greene, A. and Stuart, J. and McConnell, R. and Bruzewicz, C. D. and Sage, J. M. and Chiaverini, J.},
  journal = {Phys. Rev. A},
  volume = {97},
  issue = {2},
  pages = {020302},
  numpages = {6},
  year = {2018},
  month = {2},
  publisher = {American Physical Society}
}

@article{wineland1998,
  title={Experimental issues in coherent quantum-state manipulation of trapped atomic ions},
  author={Wineland, David J and Monroe, Christopher and Itano, Wayne M and Leibfried, Dietrich and King, Brian E and Meekhof, Dawn M},
  journal={Journal of research of the National Institute of Standards and Technology},
  volume={103},
  number={3},
  pages={259},
  year={1998}
}

@misc{Biener2022-3D-trap,
    title = {{Miniature ion traps for fast, high-fidelity and scalable quantum computations}},
    year = {2022},
    author = {Biener, Juergen and Haeffner, Hartmut and Matthiesen, Clemens and Magidash, Eli and Oakdale, James Spencer and Xia, Xiaoxing},
    number = {WO2022046306A2},
    month = {3},
    url = {https://patents.google.com/patent/WO2022046306A2/en?inventor=hartmut+haeffner&oq=hartmut+haeffner},
    keywords = {electrode, electrodes, forming, ion trap, substrate}
}

@misc{Beverland-2022-resources-scaling-QC,
    title = {{Assessing requirements to scale to practical quantum advantage}},
    year = {2022},
    author = {Beverland, Michael E. and Murali, Prakash and Troyer, Matthias and Svore, Krysta M. and Hoefler, Torsten and Kliuchnikov, Vadym and Low, Guang Hao and Soeken, Mathias and Sundaram, Aarthi and Vaschillo, Alexander},
    month = {11},
    url = {https://arxiv.org/abs/2211.07629v1},
    arxivId = {2211.07629}
}

@article{Mehta2016,
author={Mehta, Karan K.
and Bruzewicz, Colin D.
and McConnell, Robert
and Ram, Rajeev J.
and Sage, Jeremy M.
and Chiaverini, John},
title={Integrated optical addressing of an ion qubit},
journal={Nature Nanotechnology},
year={2016},
month={12},
day={01},
volume={11},
number={12},
pages={1066-1070}
}

@article{Xu2025,
author={Xu, Shuqi
and Xia, Xiaoxing
and Yu, Qian
and Parakh, Abhinav
and Khan, Sumanta
and Megidish, Eli
and You, Bingran
and Hemmerling, Boerge
and Jayich, Andrew
and Beck, Kristin
and Biener, Juergen
and H{\"a}ffner, Hartmut},
title={3D-printed micro ion trap technology for quantum information applications},
journal={Nature},
year={2025},
month={9},
day={03}
}

@article{Yu2022,
  title = {Feasibility study of quantum computing using trapped electrons},
  author = {Yu, Qian and Alonso, Alberto M. and Caminiti, Jackie and Beck, Kristin M. and Sutherland, R. Tyler and Leibfried, Dietrich and Rodriguez, Kayla J. and Dhital, Madhav and Hemmerling, Boerge and H\"affner, Hartmut},
  journal = {Phys. Rev. A},
  volume = {105},
  issue = {2},
  pages = {022420},
  numpages = {10},
  year = {2022},
  month = {2},
  publisher = {American Physical Society}
}

@article{Brownnutt2015,
  title = {Ion-trap measurements of electric-field noise near surfaces},
  author = {Brownnutt, M. and Kumph, M. and Rabl, P. and Blatt, R.},
  journal = {Rev. Mod. Phys.},
  volume = {87},
  issue = {4},
  pages = {1419--1482},
  numpages = {64},
  year = {2015},
  month = {Dec},
  publisher = {American Physical Society}
}

@article{Wright2013,
year = {2013},
month = {3},
publisher = {IOP Publishing},
volume = {15},
number = {3},
pages = {033004},
author = {Wright, Kenneth and Amini, Jason M and Faircloth, Daniel L and Volin, Curtis and Charles Doret, S and Hayden, Harley and Pai, C-S and Landgren, David W and Denison, Douglas and Killian, Tyler and Slusher, Richart E and Harter, Alexa W},
title = {Reliable transport through a microfabricated X-junction surface-electrode ion trap},
journal = {New Journal of Physics}
}

\end{document}